\documentstyle[preprint,epsf,aps]{revtex}
\begin{document}
\title{Producing the {\it a priori} pure entangled states by type II prametric downconversion}
\author{Wang Xiang-bin\thanks{email: wang$@$qci.jst.go.jp} 
\\
        Imai Quantum Computation and Information project, ERATO, Japan Sci. and Tech. Corp.\\
Daini Hongo White Bldg. 201, 5-28-3, Hongo, Bunkyo, Tokyo 113-0033, Japan}

\maketitle 
\begin{abstract}
We propose a scheme to produce the pure entangled states through  type II downconversion.
In the scheme, the vacuum states are excluded and the a prori pure entangled states are produced and 
verified without destroying the state itself. This can help to carry out the many unditional experiments related
to quantum entanglement.  
\end{abstract}
The resource of quantum entangled entangled states play a fundamentally role in testing the quantum laws
related to the nolocality and in many application in quantum information such as the quantum teleportation\cite{bennett},
quantum key distribution\cite{ekert} and quantum computation\cite{chuang}.
Many applications  of the quantum entanglement properties, e.g., the unconditional quantum 
teleportation\cite{kimble0,braun} require the preshared entangled states. 
So far the parametric downconversion\cite{kwiat} seems to be the main way to yield the polarized entangled
state
\begin{eqnarray}
|\Psi^-\rangle =\frac{1}{\sqrt 2}(|H\rangle|V\rangle-|V\rangle|H\rangle)
.\end{eqnarray}
However, in the downconversion process, actually most of the times it yields the vacuum. There is only
a small probability to yield the anti-symmetric state defined above.Due to this fact, many experiments
so far have been only carried out by the postselection\cite{bou,bou,pant}. The best way to 
overcome this draw back is to first produce many copies of any one of the {\it a priori} known  pure Bell states as the following 
\begin{eqnarray}
|\Psi^{\pm}\rangle =\frac{1}{\sqrt 2}(|H\rangle|V\rangle\pm|V\rangle|H\rangle)
\end{eqnarray}
and
\begin{eqnarray}
|\Phi^{\pm}\rangle =\frac{1}{\sqrt 2}(|H\rangle|H\rangle\pm|V\rangle|V\rangle)
\end{eqnarray}
without destroying measurement and then use the pure entangled states to carry out the quantum task.
Here we show that we can in principle obtain the {\it a prori} pure entangled states by appropriately
 implementating
 the well known properties polarizing beam splitters.
 
Let us first consider the well known properties of a polarizing beam splitter(PBS). As it was shown
in Fig. 1, a PBS reflects  vertical polarized photons and transmits the horizontally polarized photons. 
 As it was shown in \cite{pana,pann}, this property can be applied to make an incomplete Bell measurement.
Cosider the case in Fig.2. Suppose there is one incident photon in each side of the PBS with their polarization
totally unknown in principle. If we can find a photon in the outcome in each side of the PBS, then the incident
photons must be both reflected or both trasmitted. That is to say, they must have the same polarization.
Therefore, once we find the fact of a photon in each side of the PBS,
 the incident beams must have collapsed to state
$|\Phi^+\rangle$ or $|\Phi^-\rangle$.  
To distinguish $\Phi^+\rangle$ and $|\Phi^-\rangle$, we may take a Hadamad transformation to each of the outcome
beams first and then let them pass aditional polarizing beam splitters and finally make a detection using 4 photon detectors(see Fig. 3).
The Hadamad transformation is:
\begin{eqnarray}
U_H=\frac{1}{\sqrt 2}\left(\begin{array}{cc}1&1\\1&-1\end{array}\right),
\end{eqnarray}     
i.e., 
\begin{eqnarray}
U_H|H\rangle=\frac{1}{\sqrt 2}(|H\rangle +|V\rangle)\label{had};
\end{eqnarray}
and 
\begin{eqnarray}
U_H|V\rangle=\frac{1}{\sqrt 2}(|H\rangle -|V\rangle)\label{had1}.
\end{eqnarray}
After this transformation, states $|\Phi^+\rangle$ and $|\Phi^-\rangle$ can then be distinguished by the observation about which
one of $D_1$ and $D_2$ is fired and which one of $D_3$ and $D_4$ is fired(see in Fig. 3) .  
\\ Now we see how to make yield the {\it a prori} pure entangled states by these properties of PBS.
We propose the scheme in Fig. 4, which is actually a modified scheme of ref\cite{pant}. However, by the experiment in that paper,
although the entanglement swapping is verified, no {\it a priori} entangled pairs can be produced and the entanglement swapping is only verified by the measurement which
destroies the swapped entangled states.

In the set up in Fig.4, one simply observe the coincidence that satisfies {\it all}
the following three conditions
\\1. One and only one detector from $D_1$ and $D_2$ is fired, 
\\2. One and only one detector from $D_3$ and $D_4$ is fired \\
All events does not satisfy any one of these conditions are excluded.
In this way, if $D_1$ and $D_3$ or $D_2$ and $D_4$ are fired, the state for beam 4 and 3 is deterministically
collapsed into the Bell state $|\Phi^+\rangle$.  If $D_1$ and $D_4$ or $D_2$ and $D_3$ are fired, then we know
that the state for beam 3 and 4 is $|\Phi^-$. Whenever a  coincidence event happens, 
we always know it exactly whether beam 3 and beam 4 are on the state $|\Phi^+\rangle$ or $|\Phi^-\rangle$.
(Note that in a real experiment, extra devices could be required to adjust the path
of beam 1 and beam 2 so that to gurantee that they reach the PBS1 simultaneously.)

 Such a setting has ruled out any possibility  that two pairs of entangled states produced in 
one side of the nolinear crystal and
nothing produced in the other side of the nolinear crystal. Suppose such a event happens. 
For simplicity we assume two pairs in the left side
of the crystal. Now beam 1 must consist of two photons and beam 2 consists of nothing.
Explicitly, the state of two
pair term for beam 1 and 4 in a type II parametric downconversion process is\cite{bounature}
\begin{eqnarray}
\frac{1}{\sqrt 3}(|2H,0V\rangle_1|0H,2V\rangle_4-|1H,1V\rangle_1|1H,1V\rangle_4+|0H,2V\rangle_1|2H,0V\rangle_4
\end{eqnarray}
where $|nH,mV\rangle_i$ indicates a state including $n$ horizontally polarized photons and $m$ vertically polarized photons
in beam $i$.
In our coincidence we  require one from $D_1,D_2$ and one from $D_3,D_4$ are fired. 
For the term of $|2H\rangle$ or $|2V\rangle$, they are always on the same side of PBS$_1$ and PBS$_0$ therefore these terms are
ruled out by the requirement of our two conditions for the coincidence.
For the term of $|1H1V\rangle$, after it reaches PBS$_1$, it will distribute in different sides of PBS$_1$. However, they
will finally be at the same side of PBS$_2$, because a PBS transmits state $|H\rangle$ and reflects state $|V\rangle$.
Therefore, whenever the event of two pairs in the same side  of the crystal happens, 
photons in the upper beam will be finally 
 at the same side of PBS$_0$  therefore
either both $D_3$ and $D_4$ will be silent or both $D_1$ and $D_2$ will be silent. 
 These unwanted events are totally excluded by our definition of the coincidence.\\
So far we have had a scheme to produce the pure entangled states. 
This scheme itself can be regarded as a  
demonstration of the non-post selection of quantum swapping. Obviously, this scheme can also be used to demonstrate
many other notrivial unconditional quantum tasks such as the unditional 
quantum teleportation\cite{bou}, since this scheme can offer us {\it a priori} pure entangled states. 

This scheme has a very high feasibility in practice. 
In the set-up, even the photon detectors with very low efficiency can work very well, which differs totally from
the cascaded method\cite{braun}.
And also, our scheme is obviously only a slight modification
of the exisiting quantum swapping experiment\cite{pant}. I believe it 
can be carried out by the exisiting technology easily.  

{\bf Acknowledgement:} I thank Prof Imai H for support. I thank Dr Matsumoto K and Tomita A for useful discussions.
I thank Dr Pan JW( U. Viena) for pointing out ref\cite{pana}. 

\begin{figure}
\begin{center}
\epsffile{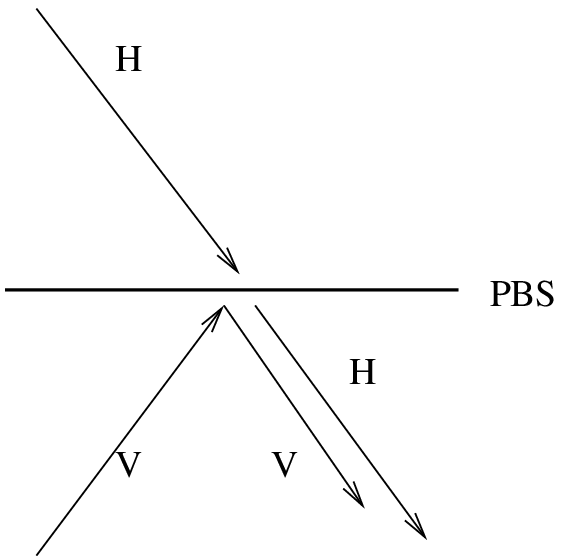}
\end{center}
\caption{ A schematic diagram for the property of a polarizing beam spliter(PBS). 
It transmits a horizontally polarized
photon $H$ and reflects a vertically polarized photon $V$.} 
\end{figure}
\newpage
\begin{figure}
\begin{center}
\epsffile{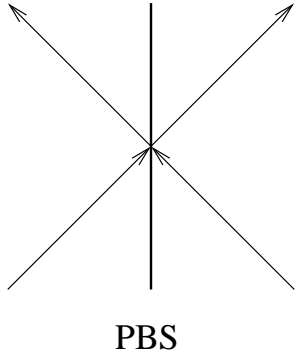}
\end{center}
\caption{ Two unkuown polarization states are incident from each side of a PBS. If there one photon in each side
of the PBS in the outcome beams, the state must be one of $|\Phi^{\pm}\rangle$.}
\end{figure}
\begin{figure}
\begin{center}
\epsffile{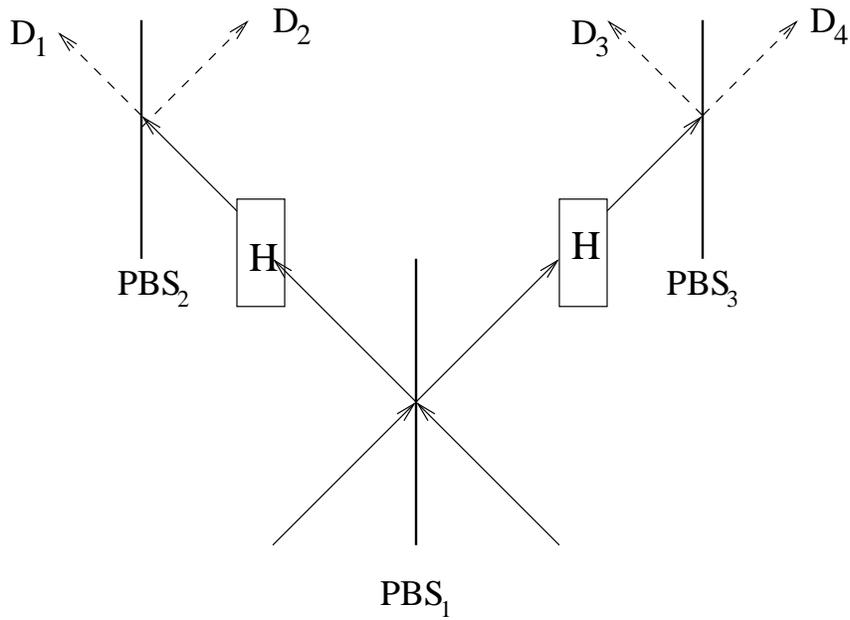}
\end{center}
\caption{An incomplete Bell measurement with polarizing beam splitters. 
H represents for a hadamad transform as defined in eq.(\ref{had},\ref{had1}). Symbol M indicates a mirrow. Simble D$_{i}$ for the
$i$'th photon detector.}
\end{figure}
\begin{figure}
\begin{center}
\epsffile{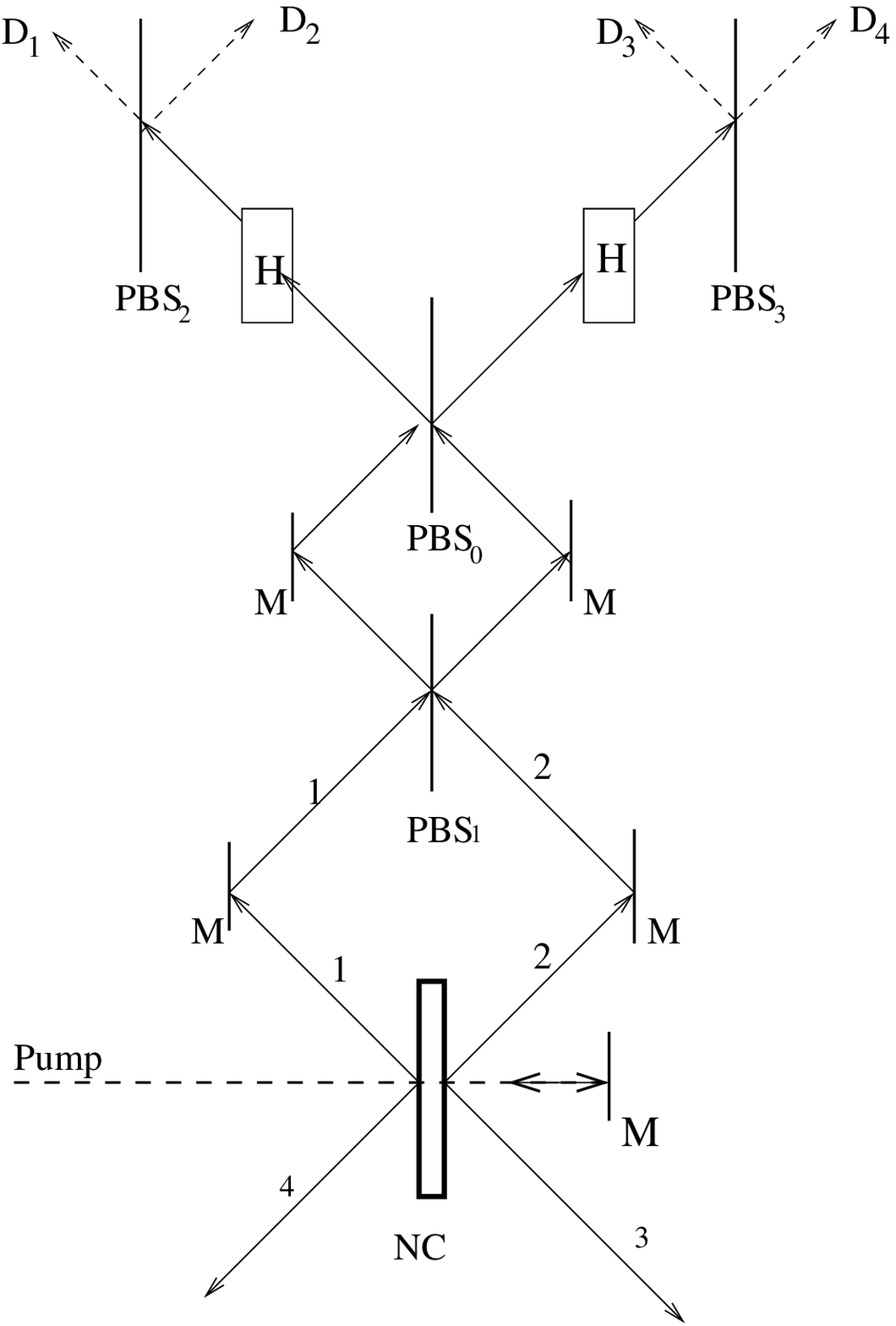}
\end{center}
\caption{Unconditional quantum teleportation with polarizing beam splitters. NC represents for the nonlinear crystal
used in the type II parametric downconversion. In a real experiment, extra devices could be
required to adjust the paths of beam 1 and beam 2 so that they reach PBS$_1$ simultaneously. 
After the meassurement D${_i}$, if we obtain a concidence, we will know the exact Bell state for beam 3 and 4, and we just store
them for the future use. Note the PBS$_0$ here plays an important role to exclude any possibility that two pairs emitted in the same
side of the nolinear crystal.}
\end{figure}

\begin{thebibliography}{99}
\addcontentsline{toc}{chapter}{Bibliography}
\bibitem{bennett} C.H. Bennette et al, Teleporting an unknown quantum state via dual
classical and EPR channels. Phys. Rev. Lett., 70:1895-1999(1993).  
\bibitem{ekert} Ekert A, Rarity J. G., Tapster P. R. and Palma G. M., Phys. Rev. Lett., 84, 4729(1992).
\bibitem{chuang} Nielson M and Chuang I.L., Quantum Computation and Quantum Information, Cambridge Press, 2000.
\bibitem{kimble0} Kimble H.J.,A posteriori teleportation, Nature 394, 841(1998).
\bibitem{braun} Kok P and Braunstein S.L., Postselected vers nonpostselected quantum teleportation using parametric down-conversion.
Phys. Rev. A 61, 042304(2000). 
\bibitem{kwiat}Kwiat P. G., Mattle K, Weinfuter H and Zeilinger A, Phys. Rev. Lett. 75, 4337(1995).
\bibitem{pana} Pan J.W. and Zeilinger A, Phys. Rev. A57, 2208. 
\bibitem{pann} Pan J W, Simon C, Brukner C and  Zeilinger A, Nature, 410, 1067(2001).
\bibitem{bou} Bowmeester D et al, Experimental quantum teleportation, Nature, 390, 6660(1997).
\bibitem{pant} Pan J.W. et al, Phys. Rev. Lett. 80, 3891{1998}.
\bibitem{bounature} Lamas-Linares A, Howell J.C. and Bouwmeester D., Stimulated emission of polarized-entangled photons.
Nature, 412, 887(2001).
\end{thebibliography}
\end{document}